\documentclass{aa}
\usepackage{graphicx}
\usepackage{natbib}
\usepackage{times}
\usepackage{fancyhdr}
\usepackage{bm,url}
\graphicspath{{./fig/}{./png/}}

\newcommand{\EQ}{\begin{equation}}
\newcommand{\EN}{\end{equation}}
\newcommand{\EQA}{\begin{eqnarray}}
\newcommand{\ENA}{\end{eqnarray}}
\newcommand{\eq}[1]{(\ref{#1})}

\newcommand{\Eq}[1]{Equation~(\ref{#1})}
\newcommand{\Eqs}[2]{Equations~(\ref{#1}) and~(\ref{#2})}

\newcommand{\Sec}[1]{Section~\ref{#1}}

\newcommand{\Fig}[1]{Figure~\ref{#1}}

\newcommand{\Tab}[1]{Table~\ref{#1}}

\newcommand{\bra}[1]{\langle #1\rangle}
\newcommand{\bbra}[1]{\left\langle #1\right\rangle}

{}
{}
{}

{}
{}
{}
{}
{}
{}
{}
{}
{}
{}
{}
{}
{}
{}
{}
{}
{}
{}
{}
{}

{}
{}
{}

{}

{}
{}

\newcommand{\yyy}{\hat{\mbox{\boldmath $y$}} {}}

\newcommand{\xx}{\bm{x}}

\newcommand{\uu}{\mbox{\boldmath $u$} {}}

\newcommand{\FF}{\mbox{\boldmath $F$} {}}

\newcommand{\GG}{\mbox{\boldmath $G$} {}}

\newcommand{\nab}{\mbox{\boldmath $\nabla$} {}}
\newcommand{\OO}{\bm{\Omega}}

\newcommand{\oo}{\mbox{\boldmath $\omega$} {}}

\newcommand{\SSSS}{\mbox{\boldmath ${\sf S}$} {}}

\newcommand{\DD}{{\rm D} {}}

\newcommand{\dd}{{\rm d} {}}

\def\la{\mathrel{\mathchoice {\vcenter{\offinterlineskip\halign{\hfil
$\displaystyle##$\hfil\cr<\cr\sim\cr}}}
{\vcenter{\offinterlineskip\halign{\hfil$\textstyle##$\hfil\cr<\cr\sim\cr}}}
{\vcenter{\offinterlineskip\halign{\hfil$\scriptstyle##$\hfil\cr<\cr\sim\cr}}}
{\vcenter{\offinterlineskip\halign{\hfil$\scriptscriptstyle##$\hfil\cr<\cr\sim\cr}}}}}
\def\ga{\mathrel{\mathchoice {\vcenter{\offinterlineskip\halign{\hfil
$\displaystyle##$\hfil\cr>\cr\sim\cr}}}
{\vcenter{\offinterlineskip\halign{\hfil$\textstyle##$\hfil\cr>\cr\sim\cr}}}
{\vcenter{\offinterlineskip\halign{\hfil$\scriptstyle##$\hfil\cr>\cr\sim\cr}}}
{\vcenter{\offinterlineskip\halign{\hfil$\scriptscriptstyle##$\hfil\cr>\cr\sim\cr}}}}}
\def\Ma{\mbox{\rm Ma}}
\def\Co{\mbox{\rm Co}}
\def\Sh{\mbox{\rm Sh}}
\def\St{\mbox{\rm St}}

\def\Rey{\mbox{\rm Re}}

\def\Co{\mbox{\rm Co}}

\def\csz{c_{\rm s0}}
\def\cs{c_{\rm s}}

\def\kf{k_{\rm f}}

\def\kom{k_{\omega}}

\def\urms{u_{\rm rms}}
\def\orms{\omega_{\rm rms}}

\def\half{{\textstyle{1\over2}}}

\def\onethird{{\textstyle{1\over3}}}

\newcommand{\s}{\,{\rm s}}

\newcommand{\cm}{\,{\rm cm}}

\newcommand{\kms}{\,{\rm km/s}}

\newcommand{\pc}{\,{\rm pc}}

\newcommand{\yaj}[3]{ #1, {AJ,} {#2}, #3}

\newcommand{\yapj}[3]{ #1, {ApJ,} {#2}, #3}

\newcommand{\yapjl}[3]{ #1, {ApJ,} {#2}, #3}

\newcommand{\yana}[3]{ #1, {A\&A,} {#2}, #3}

\newcommand{\ypf}[3]{ #1, {Phys.\ Fluids,} {#2}, #3}

\newcommand{\yaraa}[3]{ #1, {ARA\&A,} {#2}, #3}

\newcommand{\yprl}[3]{ #1, {Phys.\ Rev.\ Lett.,} {#2}, #3}

\newcommand{\ymn}[3]{ #1, {MNRAS,} {#2}, #3}

\newcommand{\yprd}[3]{ #1, {Phys.\ Rev.\ D,} {#2}, #3}
\newcommand{\ypre}[3]{ #1, {Phys.\ Rev.\ E,} {#2}, #3}

\newcommand{\yjour}[4]{ #1, {#2}, {#3}, #4}

\newcommand{\ybook}[3]{ #1, {#2} (#3)}
\newcommand{\yproc}[5]{ #1, in {#3}, ed.\ #4 (#5), #2}

\begin{document}

\titlerunning{Vorticity production through rotation, shear, and baroclinicity}
\authorrunning{}

\title{Vorticity production through rotation, shear, and baroclinicity}
\author{F. Del Sordo\inst{1,2} \and A. Brandenburg\inst{1,2}}
\institute{Nordita, AlbaNova University Center, Roslagstullsbacken 23,
SE-10691 Stockholm, Sweden
\and Department of Astronomy, AlbaNova University Center,
Stockholm University, SE 10691 Stockholm, Sweden}

\date{\today,~ $ $Revision: 1.93 $ $}

\abstract{
In the absence of rotation and shear, and under the assumption of
constant temperature or specific entropy, purely potential forcing
by localized expansion waves is known to produce irrotational flows
that have no vorticity.
}{%
Here we study the production of vorticity under idealized conditions
when there is rotation, shear, or baroclinicity, to address the problem
of vorticity generation in the interstellar medium in a systematic fashion.
}{%
We use three-dimensional periodic box numerical simulations to
investigate the various effects in isolation.
}{%
We find that for slow rotation, vorticity production
in an isothermal gas is small in the sense that the ratio
of the root-mean-square values of vorticity and velocity
is small compared with the wavenumber of the energy-carrying motions.
For Coriolis numbers above a certain level, vorticity production saturates at a
value where the aforementioned ratio becomes comparable with the
wavenumber of the energy-carrying motions.
Shear also raises the vorticity production, but no saturation is found.
When the assumption of isothermality is dropped, there is significant
vorticity production by the baroclinic term once the turbulence becomes
supersonic.
In galaxies, shear and rotation are estimated to be insufficient to
produce significant amounts of vorticity, leaving therefore only the
baroclinic term as the most favorable candidate.
We also demonstrate vorticity production visually as a result of colliding shock
fronts.
}{}


\maketitle

\section{Introduction}

Turbulence in the interstellar medium (ISM) is believed to be driven by
supernova explosions.
Such events inject sufficient amounts of energy to sustain turbulence with
rms velocities of $\sim10\kms$ and correlation lengths of up to $100\pc$
\citep{Beck96}.
Simulations of such events can be computationally quite demanding,
because the bulk motions tend to be supersonic and the flows involve
strong shocks in the vicinity of individual explosion sites,
as was seen early on in two-dimensional simulations \citep{RB95}.
Nevertheless, such simulations are able to reproduce a number of physical phenomena
such as the observed volume fractions of hot, warm, and cold gas
\citep{RBK96,K99}, the statistics of pressure fluctuations \citep{MacLow},
the effects of the magnetic field \citep{dAB05},
and even dynamo action \citep{G08,Giss09,Hanasz09}.
These simulations tend to show the development of significant amounts
of vorticity, which is at first glance surprising.
Indeed, each supernova drives the gas radially outward and can roughly
be described by radial expansion waves.
In such a description, turbulence is forced by the gradient of a potential
that consists of a time-dependent spherical blob at random locations.
Obviously, such a forcing is irrotational, so no vorticity is produced.

Earlier work of \cite{MB06} showed that under isothermal conditions only
the viscous force can produce vorticity and that this becomes negligible
in the limit of large Reynolds numbers or small viscosity.
In principle, vorticity can also be amplified akin to the dynamo effect
by the $\nab\times(\uu\times\oo)$ term, which is analogous to the
induction term in dynamo theory, where $\oo$ plays the role of the
magnetic field.
However, neither this nor the viscosity effect were found to operate
-- even at numerical resolutions of up to $512^3$ meshpoints.
This disagreed with subsequent simulations by \cite{Federrath}, who solved
the isothermal inviscid Euler equations with irrotational forcing using
the {\sc Flash Code}.
They found significant vorticity generation in proximity to shocks where
some kind of effective numerical viscosity must have acted.

Given that under isothermal conditions, only viscosity can lead to
vorticity production, one must ask whether numerical viscosity or
effective viscosity needed to stabilize numerical codes might have contributed
to the production of vorticity in some of the earlier works.
Indeed, it is possible that the directional operator splitting used in
the {\sc Flash Code} may have been responsible for spurious vorticity
generation in the work of \cite{Federrath}; (R. Rosner, private communication).
On the other hand, when cooling and heating functions are included
to perform more realistic simulations of the ISM, vorticity could
be produced by the baroclinic term.
Furthermore, even in the isothermal case, in which the baroclinic term
vanishes, vorticity could be produced if there is rotation and/or shear.

The baroclinic term results from taking the curl of the pressure gradient
term and is proportional to the cross product of the gradients of pressure and density.
This term can play an important role when the assumptions of isothermality
or adiabaticity are relaxed.
Indeed, the baroclinic term can also be written as
the cross product of the gradients of entropy and temperature.
This formulation highlights the need for non-ideal effects,
because in the absence of any other heating or cooling mechanisms, the
entropy is just driven by viscosity.
Again, it is not obvious that in the absence of additional heating and
cooling much vorticity can be produced.
On the other hand, it is clear that viscous heating must be significant
even in the limit of vanishing viscosity, because the velocity gradients
can be very large, especially in shocks.
Of course, the assumption about additional heating and cooling is not
realistic for the interstellar medium and will need to be relaxed.
Finally, there are the effects of rotation and shear, that can contribute
to the production of vorticity even in the absence of baroclinicity.

The goal of this paper is to study the relative importance of the individual
effects that contribute to vorticity production.
It is then advantageous to restrict oneself to simplifying conditions that
allow one to identify the governing effects.
An important simplification is the restriction to weakly supersonic
conditions so that shocks and other sharp structures can still be
resolved with just a uniform and constant viscosity.
We also neglect the effects of stratification which can only indirectly
contribute to vorticity production.
In fact, a constant gravitational acceleration drops out when taking the curl.
Only in the non-isothermal and non-isentropic case can gravity contribute
to vorticity production by enhancing the effect of the baroclinic term.
We begin with a preliminary discussion and a qualitative analysis of the
important terms in the vorticity equation.

\section{Preliminary considerations}

We recall that in the absence of baroclinicity, rotation, and shear, the
curl of the evolution equation of the velocity is given by
\citep[see, e.g.,][]{MB06}
\EQ
{\partial\oo\over\partial t}=\nab\times(\uu\times\oo-\nu\nab\times\oo)
+\nu\nab\times\GG,
\label{dodt}
\EN
where $\nu$ is the kinematic viscosity (assumed constant) and
$G_i=2{\sf S}_{ij}\nabla_j\ln\rho$ is a part of the viscous force
that has non-vanishing curl even when the flow is purely irrotational.
Here,
\EQ
{\sf S}_{ij}=\half(u_{i,j}+u_{j,i})-\onethird\delta_{ij}u_{k,k}
\label{StrainMatrix}
\EN
is the traceless rate of strain matrix, and commas denote partial
differentiation.
The $\GG$ term breaks the formal analogy with the induction equation.
It is convenient to express the resulting rms vorticity in terms of
the typical wavenumber $\kom$ of vortical structures which we define as
\EQ
\kom=\orms/\urms.
\EN
We monitor the ratio $\kom/\kf$, where $\kf$ is the
adopted nominal forcing wavenumber.
In \cite{MB06}, the resulting vorticity, expressed in terms of
the ratio $\kom/\kf$, was found to be zero within error bars.
This result is compatible with the idea that the $\nu\nab\times\GG$
term in \Eq{dodt} is insignificant for vorticity production.
By contrast, in vortical turbulence and at moderate values of
the Reynolds number, $\kom/\kf$ is found to be of the order of unity
\citep{B01}, although one should expect a mild increase proportional
to the square root of the Reynolds number as this number increases.

\subsection{Rotation}

Rotation leads to the addition of the Coriolis force, $2\OO\times\uu$,
in the evolution equation for the velocity.
Taking the curl, we obtain the vorticity equation \eq{dodt} with
two additional terms, both proportional to $\OO$, so we have
\EQ
{\partial\oo\over\partial t}
=...-2\OO\nab_\perp\cdot\uu_\perp+2\OO\cdot\nab\uu_\perp,
\label{Coriolis}
\EN
where the dots denote the other terms in \Eq{dodt} that we discussed already.
In order to estimate the production of vorticity, one could
derive an evolution equation for the enstrophy density,
$\half\oo^2$, by multiplying the right-hand
side of \Eqs{dodt}{Coriolis} by $\oo$, and use a closure assumption
for the resulting triple correlations.
However, it is then difficult to obtain a useful prediction for
$\omega_{\rm rms}$, because the right-hand side of such an equation
would necessarily be proportional to $\oo$ and would
therefore vanish, unless $\omega_{\rm rms}$ was different from zero
to begin with.
Instead, we estimate $\omega_{\rm rms}$ by computing the rms value of
$\partial\oo/\partial t$ and replacing it by  $\omega_{\rm rms}/\tau_\Omega$,
where $\tau_\Omega$ is a typical time scale of the problem.
This leads to
\begin{equation}
\orms\approx2\Omega\tau_\Omega\bbra{(\nab_\perp\cdot\uu_\perp)^2
+(\nabla_\parallel\uu_\perp)^2}^{1/2},
\label{ormsROT}
\end{equation}
where $\nab_\perp$ and $\nabla_\parallel$ denote derivatives in the
directions perpendicular and parallel to the rotation axis and $\uu_\perp$
is the velocity vector perpendicular to the rotation axis.
Using Cartesian coordinates where $\OO$ points in the $z$ direction, we have
\begin{equation}
\orms\approx2\Omega\tau_\Omega\bbra{(u_{x,x}+u_{y,y})^2
+u_{x,z}^2+u_{y,z}^2}^{1/2}.
\label{DerivativeTerms}
\end{equation}
We expect $\tau_\Omega$ to be comparable to the turnover time,
$\tau=(\urms\kf)^{-1}$.
We expect the rms values of the velocity derivative term in
\Eq{DerivativeTerms} to be comparable to the rms velocity and some
inverse length scale.
Typically, one would expect it to be proportional to $\urms\kf$, although,
again, there can be an additional dependence on the square root of the
Reynolds number.
However, for fixed Reynolds number, and not too rapid rotation, we expect
$\orms$ to increase linearly with the Coriolis number, i.e.,
\EQ
\Co=2\Omega\tau,\quad\mbox{where}\quad\tau=(\urms\kf)^{-1}.
\EN
Thus, we expect $\kom/\kf=\St_\Omega\,\Co$, where we have defined
an effective rotational Strouhal number,
\EQ
\St_\Omega=\tau_\Omega^{\rm eff}\urms\kf.
\label{St_Om}
\EN
We regard this as a fit parameter that will emerge as a result of the
simulations.
We have here introduced the quantity $\tau_\Omega^{\rm eff}$, where
$\tau_\Omega^{\rm eff}/\tau_\Omega$ is given by the ratio of
the velocity gradient terms divided by $\urms\kf$.
However, for larger values of $\Co$ there may be a departure from a
linear dependence between $\kom/\kf$ and $\Co$.
(We note that, apart from a possible $4\pi$ factor, the Coriolis number
is just the inverse Rossby number.)
One aim of this paper is therefore to verify this dependence
from simulations and to determine empirically the value of $\tau_\Omega$.

\subsection{Shear}
\label{Shear}

In the presence of linear shear with $\uu^S=(0,Sx,0)$, the
evolution equation for the departure from the mean shear attains
additional terms, $-\uu^S\nab\cdot\uu-\uu\cdot\nab\uu^S$.
This implies a dependence of $\orms$ on $S$,
analogous to the $\Omega$ dependence discussed above.
In components form, this means that
\begin{equation}
\orms\approx S\tau_S\bbra{(u_{x,x}+u_{y,y})^2
+u_{x,z}^2+u_{z,y}^2+O(xu'')}^{1/2},
\label{DerivTermS}
\end{equation}
which is quite similar to \Eq{ormsROT}, except that in the penultimate term
in angular brackets the indices are now interchanged, i.e.\ we now have
$u_{z,y}$ instead of $u_{y,z}$.
In analogy to $\tau_\Omega$, we define $\tau_S$ as a typical time scale
of the problem and we expect it to
be again related to the turnover time $\tau$.
The $O(xu'')$ denotes the presence of additional terms that are
proportional to $x$ and to second derivatives of $\uu$.
However, when adopting the shearing box approximation with
shearing-periodic boundaries \citep{GL65,WT88}, each point in the $xy$ plane
is statistically equivalent.
We would therefore not expect there to be a systematic $x$ dependence,
which corresponds to the assumption of Galilean invariance that is
sometimes used in the study of turbulent transport coefficients in
linear shear flows \citep{SS09}.
We will postpone the possibility of additional terms until later.
Since we expect $\tau_S$ to be comparable to $\tau=(\urms\kf)^{-1}$,
the rms vorticity should be proportional to the shear parameter,
\EQ
\Sh=S\tau\equiv S/\urms\kf,
\EN
although for large values of $|\Sh|$ we may expect departures from
a linear dependence.
Determining this dependence is another aim of this paper.
Again, a linear dependence is characterized by the values of $\tau_S$
and $\tau_S^{\rm eff}$, where, in analogy with the previous case with
rotation, the ratio $\tau_S^{\rm eff}/\tau_S$ is given by the derivative
term in \Eq{DerivTermS}, normalized by $\urms\kf$.
A convenient non-dimensional measure of the value of $\tau_S^{\rm eff}$
is what we call the shear Strouhal number,
\EQ
\St_S=\tau_S^{\rm eff}\urms\kf,
\label{St_Om2}
\EN
which can be determined provided there is a range in $\Sh$ over which
$\omega_{\rm rms}$ increases linearly with $\Sh$.

The study of vorticity production by rotation and shear is quite
independent of thermodynamics and can in principle be studied even
in the incompressible case.
However, in the present paper we study this effect in the weakly
compressible case of low Mach numbers and under the assumption of an
isothermal equation of state, where the baroclinic term vanishes.

\subsection{Baroclinicity}

As mentioned in the introduction, the baroclinic term, proportional
to $\nab\rho\times\nab p$, emerges when taking the curl of the pressure
gradient term, $\rho^{-1}\nab p$.
This term can also be written as
\EQ
\rho^{-1}\nab p=\nab h-T\nab s,
\EN
where $h$ and $s$ are specific enthalpy and specific entropy,
respectively, and $T$ is the temperature.
Thus, we have
\EQ
{\partial\oo\over\partial t}=...+\nab T\times\nab s.
\label{dodt_baro}
\EN
In order to study the effect of the baroclinic term, it is useful to
look at the dependence of the mean angle $\theta$ between the gradients
of $s$ and $T$, defined via
\EQ
\sin^2\!\theta=\bra{(\nab T\times\nab s)^2}/\bra{(\nab T)^2}\bra{(\nab s)^2}.
\EN
An important aspect is then to study first the dependence of the
rms values of the gradients of $s$ and $T$.
We can do this by looking at a one-dimensional model where, of course,
$\theta=0$.

Next, we need to determine $\theta$ from three-dimensional simulations.
The hope is then that we can express baroclinic vorticity production in
the form
\EQ
\kom/\kf=\St_{\rm baro}(\nab T)_{\rm rms}(\nab s)_{\rm rms}\sin\theta/\urms^2\kf^2.
\EN
On dimensional grounds we expect the product of $(\nab T)_{\rm rms}$
and $(\nab s)_{\rm rms}$ to be of the order of $\urms^2\kf^2$, and so
a possible ansatz would be
\EQ
\kom/\kf=\St_{\rm baro}^{\rm eff}\sin\theta,
\EN
where we have subsumed the scalings of $(\nab T)_{\rm rms}$ and
$(\nab s)_{\rm rms}$ in that of an effective baroclinic Strouhal
number $\St_{\rm baro}^{\rm eff}$.

An important issue is the fact that viscous heating leads to a continuous
increases of the temperature.
As a result, the sound speed changes and it becomes then impossible
to study the behavior of the system in a steady state.
In order to avoid this inconvenience, we add a volume cooling term
that is non-vanishing when the local sound speed $\cs$ is different
from a given target value, $\csz$.
Thus, in the presence of finite thermal diffusivity $\chi$, and with a
cooling term governed by a cooling time $\tau_{\rm cool}$, our entropy
equation takes the form
\EQ
T{\DD s\over\DD t}=2\nu\SSSS^2+\rho^{-1}\nab\cdot(c_p\rho\chi\nab T)
-{1\over\tau_{\rm cool}}(\cs^2-\csz^2),
\label{dsdt}
\EN
where $\cs$ is the adiabatic sound speed.
We assume a perfect gas so that $\cs^2=(\gamma-1)c_p T$,
where $\gamma=c_p/c_v=5/3$ for a monatomic gas,
and $c_p$ and $c_v$ are the specific heats at constant pressure
and constant volume, respectively.
The value of $\tau_{\rm cool}$ can have an influence on the
results, so we need to consider different values.
We express $\tau_{\rm cool}$ in terms of $\csz$ and $\kf$, and define
the nondimensional quantity $\St_{\rm cool}=\tau_{\rm cool}\csz\kf$.

\section{The model}

In this paper we solve the continuity equation for the density $\rho$,
\EQ
{\DD\ln\rho\over\DD t}=-\nab\cdot\uu,
\EN
together with the momentum equation for the velocity $\uu$,
\EQ
{\DD\uu\over\DD t}=-\rho^{-1}\nab p-2\OO\times\uu-Su_x\yyy
+\nab\phi+\FF_{\rm visc},
\EN
where $\DD/\DD t=\partial/\partial t+(\uu+\uu^S)\cdot\nab$ is the
advection operator with respect to the sum of turbulent flow $\uu$ and
laminar shear flow $\uu^S$, $p$ is the pressure, $\phi$ is the forcing
potential, and
\EQ
\FF_{\rm visc}=\rho^{-1}\nab\cdot(2\nu\rho\SSSS)
\EN
is the viscous force, where $\SSSS$ was defined in \Eq{StrainMatrix}.
The forcing potential is given by
\EQ
  \phi(\xx,t)=\phi_0\ N\exp\left\{-[\xx-\xx_{\rm f}(t)]^2/R^2\right\},
\EN
where $\xx=(x,y,z)$ is the position vector,
$\xx_{\rm f}(t)$ is the random forcing position that changes abruptly
after a time interval $\Delta{}t$,
$R$ is the radius of the Gaussian, and $N$ is a non-dimensional factor
proportional to $\Delta{}t^{-1/2}$.
This ensures that the amplitude of the correlation function of $\phi$
is independent of $\Delta{}t$.
Thus, we choose $N=\sqrt{R/c_{\rm s0}\Delta{}t}$.
Since $N$ is non-dimensional,
the prefactor $\phi_0$ has the same dimension as $\phi$,
which is that of velocity squared.
We consider two forms for the time dependence of $\xx_{\rm f}$.
First, we take $\xx_{\rm f}$ such that the forcing is
$\delta$-correlated in time.
In that case, $\Delta{}t$ is equal to the length of the time step $\delta t$.
Alternatively, we choose a finite forcing time
${\delta{}t}_{\rm force}$ that defines the interval during which $\xx_{\rm f}$
remains constant, after which the forcing changes again abruptly.
Thus,
\EQ
\Delta{}t=\max\left(\delta{}t,\delta{}t_{\rm force}\right)
\EN
is equal to $\delta t$ in the $\delta$-correlated case or equal to
$\delta{}t_{\rm force}$ in the case of finite correlation time.

The work of \cite{MB06} showed that the peak of the energy spectrum
depends on the radius $R$ of the Gaussian.
Indeed, the Fourier transform of $\exp(-r^2/R^2)$ is also a Gaussian
with $\exp(-k^2/\kf^2)$, where
\EQ
\kf=2/R.
\EN
In the following we use this as our definition of $\kf$ and check
a posteriori that this is close to the position of the
peak of the ener\-gy spectrum.
In the following, we characterize our simulations in terms of the
ratio $\kf/k_1$, and consider values between 2 and 10.

We use the \textsc{Pencil Code},\footnote{
\url{http://pencil-code.googlecode.com/}} which is a non-conservative,
high-order, finite-difference code (sixth order in space and
third order in time) for solving the compressible hydrodynamic and
hydromagnetic equations.
We adopt non-dimensional variables by measuring
speed in units of a reference sound speed, $c_{\rm s0}$, and length in units of
$1/k_1$, where $k_1$ is the smallest wavenumber in the periodic domain.
This implies that the nondimensional size of the domain is $(2\pi)^3$.

In order to study the effects of rotation and shear, we ignore entropy
effects and restrict ourselves to an isothermal equation of state
with constant sound speed $\cs$.
This means that $\rho^{-1}\nab p$ reduces to $\cs^2\nab\ln\rho=\nab h$, which has
vanishing curl.
Here, $h=\cs^2\ln\rho$ is the relevant enthalpy in the isothermal case.
On the other hand, in order to study the effects of baroclinicity, we do need
to allow the entropy to vary, so we also need to solve \Eq{dsdt}, and
study the dependence of $k_\omega/\kf$ on the Mach number,
\EQ
\Ma=\urms/\cs.
\EN
In order to characterize the degree of turbulence, we define the
Reynolds number based on the energy-carrying scale, corresponding
to the typical wavenumber where the spectrum peaks, i.e.\
\EQ
\Rey=\urms/\nu\kf.
\EN
For vortical turbulence, this definition is known to be a good measure of the
ratio of the resulting turbulent viscosity divided by the molecular
diffusivity \citep{YBR03}.
The two numbers, $\Ma$ and $\Rey$, can be varied by changing $\nu$
and/or the strength of the forcing.
In all cases we use $\chi=\nu$.
Another input parameter is the forcing Strouhal number
\EQ
\St_{\rm force}=\tau_{\rm force}\urms\kf,
\EN
which is zero for $\delta$-correlated forcing and equal to about
0.3 in cases with finite correlation time.
These are also the values used by \cite{MB06}.

In the following we also consider kinetic energy and enstrophy spectra,
$E_{\rm K}(k)$ and $E_\omega(k)$, respectively.
They are normalized such that \citep{Les90}
\EQ
\int E_{\rm K}(k)\,\dd k=\half\bra{\uu^2},\quad
\int E_\omega(k)\,\dd k=\half\bra{\oo^2},
\EN
where $\half\bra{\uu^2}$ and $\half\bra{\oo^2}$ are kinetic energy
and enstrophy, respectively.
For comparison we also consider spectra of enthalpy,
$E_{\rm h}(k)$, which are normalized such that
$\int E_{\rm h}(k)\,\dd k=\half\bra{h^2}$.

Throughout this paper we assume periodic boundary conditions, except
that in the presence of shear we employ shearing-periodic boundary
conditions where the $x$ direction is periodic with respect to
positions in $y$ that shift with time, i.e.\
\EQ
f(-\half L_x,y,z,t)=f(\half L_x,y+L_xSt,z,t),
\EN
where $f$ represents any one of our four dependent variables $(\uu,\rho)$.
This boundary condition was first proposed by \cite{GL65} and has
been routinely used in local simulations of accretion disk turbulence
\citep{HGB95}.
Note, however, that
recent work of \cite{RU08} and \cite{Bodo08} called attention to the
possibility of problems with the shearing sheet approximation when the
size of the perturbations is large.
In somewhat weaker form, this problem also applies to a non-shearing
periodic box.
Indeed, we shall kept this in mind when interpreting some of the
results presented below.

\begin{figure}[t!]\begin{center}
\includegraphics[width=\columnwidth]{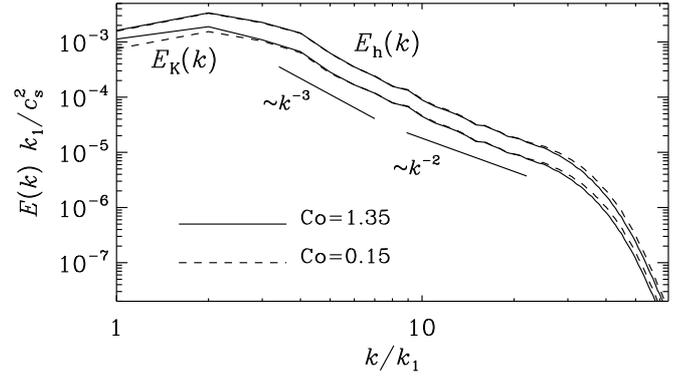}
\end{center}\caption[]{
Time-averaged kinetic energy and enthalpy spectra for two values of
the Coriolis number for $\Rey=25$ and $\St_{\rm force}=0.4$.
The two straight lines give the slopes $-2$ and $-3$, respectively.
In both cases we have $\kf/k_1=4$.
}\label{pspec_comp}\end{figure}

\begin{figure}[t!]\begin{center}
\includegraphics[width=\columnwidth]{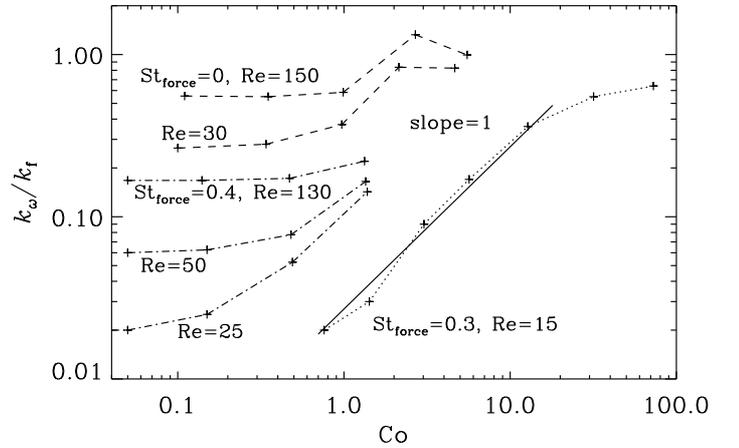}
\end{center}\caption[]{
Dependence of $\kom/\kf$ on $\Co$ for three groups of runs:
group~1 with $\Rey=15$, $\kf/k_1=10$, $\St_{\rm force}=0.3$;
group~2 with $\Rey$ between 25 and 130, $\kf/k_1=4$, $\St_{\rm force}=0.4$;
and group~3 with $\Rey=30$ and 150, $\kf/k_1=2$, $\St_{\rm force}=0$.
}\label{table_nu2em4_R02}\end{figure}

\section{Results}

We begin by studying the effect of rotation.
In \Fig{pspec_comp} we plot time-averaged
kinetic energy and enthalpy spectra,
$E_{\rm K}(k)$ and $E_{\rm h}(k)$, respectively.
Note that rotation has a tendency to move the peak of $E_{\rm K}(k)$ to the
left of the nominal value of $\kf$.
However, at the Reynolds number of 25 shown here, there is no inertial
range, but in all cases, the energy spectra show a clear viscous
dissipation range, suggesting that these runs are sufficiently well
resolved.
At somewhat larger Reynolds number or smaller forcing wavenumber,
earlier work of \cite{MB06} began to show a short $k^{-2}$ subrange.
Such a slope is predicted for shock turbulence \citep{KP73}, and it has
also been seen in the irrotational component of transonic turbulence
\citep{Porter98}.

In \Fig{table_nu2em4_R02} we plot the dependence of $\kom/\kf$ on $\Co$
for three groups of runs:
group~1 with $\Rey=15$, $\kf/k_1=10$, $\St_{\rm force}=0.3$;
group~2 with $\Rey$ between 25 and 130, $\kf/k_1=4$, $\St_{\rm force}=0.4$;
and group~3 with $\Rey=30$ and 150, $\kf/k_1=2$, $\St_{\rm force}=0$.
In all cases we use $128^3$ mesh points, average the results over
at least 200 turnover times and, in some cases, even several thousand
turnover times.
It turns out that for $\St_{\rm force}\neq0$ a linear relationship between
$\kom/\kf$ and $\Co$ is a good approximation for $\Co\la10$, where
$\kom/\kf\approx0.03\,\Co$, i.e.\ $\St_\Omega=0.03$.
Furthermore, we see from \Tab{VelDerivs} that the normalized
velocity derivative terms are all about 0.5, so the root of the sum
of their squares is slightly larger than unity, corresponding to
$\tau_\Omega^{\rm eff}/\tau_\Omega\approx1.3$.
For $\Co>10$ the value of $\kom/\kf$ seems to saturate at about unity.

\begin{figure}[t!]\begin{center}
\includegraphics[width=\columnwidth]{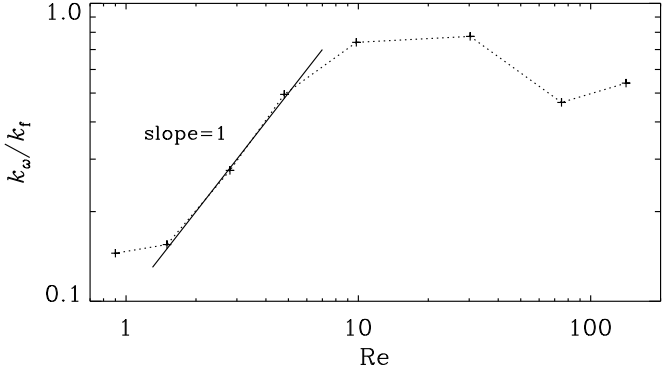}
\end{center}\caption[]{
Dependence of $\kom/\kf$ on $\Rey$ for $\Co\approx1$, $\kf/k_1=2$,
and $\St_{\rm force}=0$ ($\delta$-correlated forcing).
}\label{ptable_Redep}\end{figure}

\begin{figure}[t!]\begin{center}
\includegraphics[width=\columnwidth]{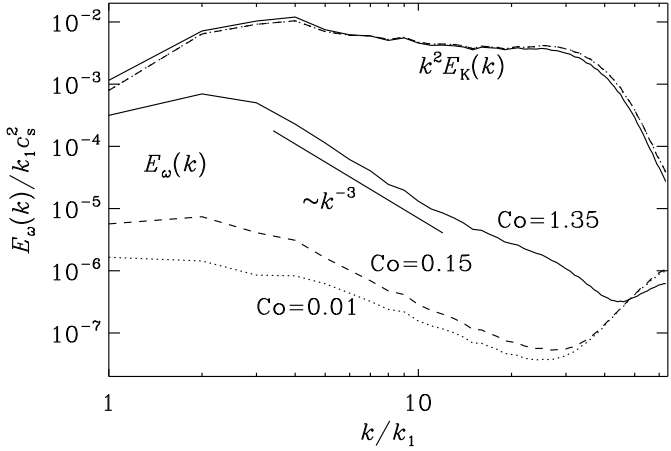}
\end{center}\caption[]{
Time-averaged enstrophy spectra, $E_\omega(k)$, compared with $k^2E_{\rm K}(k)$,
for $\Rey=25$, $\St_{\rm force}=0.4$, and three values of the Coriolis number.
The curves of $k^2E_{\rm K}(k)$ are close together and overlap for
$\Co=0.01$ (dotted) and $0.15$ (dashed), so it becomes a single
dash-dotted line.
The $k^{-3}$ slope is shown for comparison.
In all three cases we have $\kf/k_1=4$.
}\label{pspec_comp_enstr}\end{figure}

\begin{table}[b!]\caption{
Root-mean-squared values of components of the velocity derivative
tensor, normalized by $\urms\kf$, as well as the three diagonal
components of the $\bra{u_iu_j}$ tensor for 4 values of $\Co$.
}\vspace{12pt}\centerline{\begin{tabular}{l|ccccccc}
$\Co$ &   0.11 &   0.35 &   0.99 &   2.80\\
\hline
$(\nab_\perp\uu_\perp)_{\rm rms}/\urms\kf$ &   1.26 &   1.20 &   1.21 &   1.04\\
$u_{x,x}^{\rm rms}/\urms\kf$ &   0.76 &   0.74 &   0.75 &   0.70\\
$u_{y,y}^{\rm rms}/\urms\kf$ &   0.79 &   0.74 &   0.75 &   0.70\\
$u_{x,z}^{\rm rms}/\urms\kf$ &   0.49 &   0.47 &   0.48 &   0.63\\
$u_{y,z}^{\rm rms}/\urms\kf$ &   0.49 &   0.47 &   0.48 &   0.63\\
$u_{z,y}^{\rm rms}/\urms\kf$ &   0.49 &   0.47 &   0.46 &   0.58\\
$\bra{u_x^2}/\urms^2$ &   0.32 &   0.33 &   0.35 &   0.42\\
$\bra{u_y^2}/\urms^2$ &   0.34 &   0.34 &   0.35 &   0.41\\
$\bra{u_z^2}/\urms^2$ &   0.35 &   0.34 &   0.29 &   0.18\\
\label{VelDerivs}\end{tabular}}\end{table}

A similar result is also found for $\St_{\rm force}=0$, except that
there remains a spurious contamination of vorticity even for small values
of $\Co$, a limit in which we expect to observe no vorticity production.
By varying the value of $\Rey$, while keeping $\Co\approx1$ fixed, we see
that $\kom/\kf$ asymptotes to zero for sufficiently small values of $\Rey$;
see \Fig{ptable_Redep}.
This suggests that there can easily be spurious vorticity generation,
possibly due to marginally sufficient resolution.
The possibility of spurious vorticity is indeed verified by
\Fig{pspec_comp_enstr}, where we compare enstrophy spectra,
$E_\omega(k)$, with $k^2E_{\rm K}(k)$.
Note that for large values of $\Co$, the enstrophy spectrum decays
like $k^{-3}$.
However, for smaller values of $\Co$ the level of enstrophy at the
mesh scale remains approximately unchanged and is thus responsible
for the spurious vorticity found above for small values of $\Co$
and not too small values of $\Rey$.
Nevertheless for larger values of $\Co$, the production of vorticity
is an obvious effect of rotation in an otherwise potential velocity
field, and it is most pronounced at large length scales, as can also
be seen in \Fig{pspec_comp_enstr}.

\begin{figure}[t!]\begin{center}
\includegraphics[width=\columnwidth]{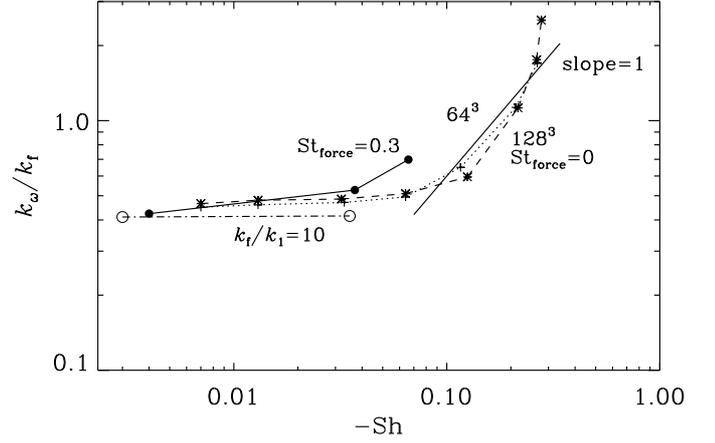}
\end{center}\caption[]{
Dependence of $\kom/\kf$ on $\Sh$ for
$\Rey\approx40$ and $\kf/k_1=2$ and $\delta$-correlated forcing
($\St_{\rm force}=0$).
Different resolutions are shown to give similar results.
At small values of $|\Sh|$, comparisons with $\St_{\rm force}=0.3$
(keeping $\kf/k_1=2$) or $\kf/k_1=10$ (keeping $\St_{\rm force}=0$)
are also shown.
}\label{table_shear}\end{figure}

Next, we study the dependence of the ratio $\kom/\kf$ on shear;
see \Fig{table_shear}.
We use a resolution of $64^3$ or $128^3$ mesh points,
average the results over at least 200 turnover times and, in cases
of lower resolution, over several thousand turnover times.
It turns out that in the presence of shear, some level of helicity
production can never be avoided -- even in the limit of small $\Sh$.
Again, this appears spurious  and demonstrates the general sensitivity
of vorticity generation on resolution effects.
An additional problem is of course the finite size of the shearing box
\citep{RU08,Bodo08}, which may be responsible for spurious vorticity
generation.
On the other hand, there is vorticity generation even for large
scale-separation ratios, $\kf/k_1=10$; see the dash-dotted line
in \Fig{table_shear}.
This suggests the possibility of a more general problem that would
not go away even in the limit of small eddies and small values of $|\Sh|$.
Nevertheless, there is a clear rise of $\kom/\kf$ when $|\Sh|>0.1$,
which is in agreement with our expectations outlined in \Sec{Shear}.
However, the slope in this relationship is rather steep, $\St_S\approx6$.
The velocity derivative terms are only slightly larger than in the case
with rotation, corresponding to $\tau_S^{\rm eff}/\tau_S\approx1.5$;
see also \Tab{VelDerivsShear}.
Tentatively, this suggests that for comparable values of $\Co$ and  $\Sh$,
$\tau_S\gg\tau_\Omega$.
On the other hand, given that even for small values of $\Sh$ there is
spurious vorticity generation, we cannot be certain that the results
are reliable for larger ones either.
The case with shear must therefore remain somewhat inconclusive.

\begin{table}[b!]\caption{
Similar to \Tab{VelDerivs}, but for the case with shear.
}\vspace{12pt}\centerline{\begin{tabular}{l|rrrrrrrr}
$\Sh$ & $-0.01$& $-0.03$& $-0.06$& $-0.12$& $-0.26$\\
\hline
$(\nab_\perp\uu_\perp)_{\rm rms}/\urms\kf$ &   1.35 &   1.36 &   1.36 &   1.37 &   0.87\\
$u_{x,x}^{\rm rms}/\urms\kf$ &   0.87 &   0.87 &   0.90 &   0.97 &   0.75\\
$u_{y,y}^{\rm rms}/\urms\kf$ &   0.81 &   0.81 &   0.76 &   0.66 &   0.47\\
$u_{x,z}^{\rm rms}/\urms\kf$ &   0.51 &   0.50 &   0.51 &   0.53 &   0.72\\
$u_{y,z}^{\rm rms}/\urms\kf$ &   0.46 &   0.47 &   0.47 &   0.46 &   0.74\\
$u_{z,y}^{\rm rms}/\urms\kf$ &   0.46 &   0.48 &   0.46 &   0.43 &   0.57\\
$\bra{u_x^2}/\urms^2$ &   0.37 &   0.38 &   0.37 &   0.36 &   0.25\\
$\bra{u_y^2}/\urms^2$ &   0.31 &   0.31 &   0.32 &   0.34 &   0.56\\
$\bra{u_z^2}/\urms^2$ &   0.32 &   0.31 &   0.31 &   0.31 &   0.25\\
\label{VelDerivsShear}\end{tabular}}\end{table}

\begin{figure}[t!]\begin{center}
\includegraphics[width=\columnwidth]{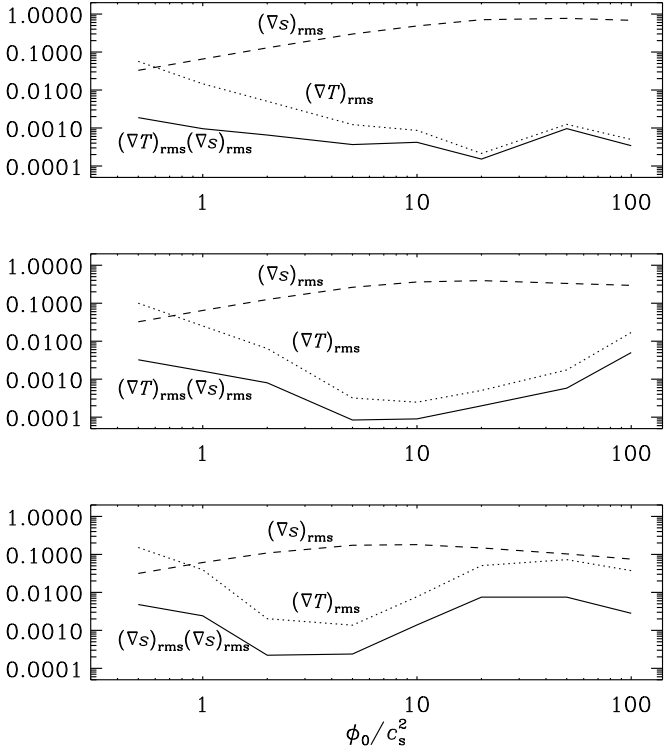}
\end{center}\caption[]{
Dependence of the rms values
of temperature and entropy on $\phi_0$ for $\nu/\cs R=1$
and $\St_{\rm cool}=0.2$ (top panel), 0.6 (middle panel), and 2 (bottom panel).
}\label{pbaro_nu4}\end{figure}

\begin{figure}[t!]\begin{center}
\includegraphics[width=\columnwidth]{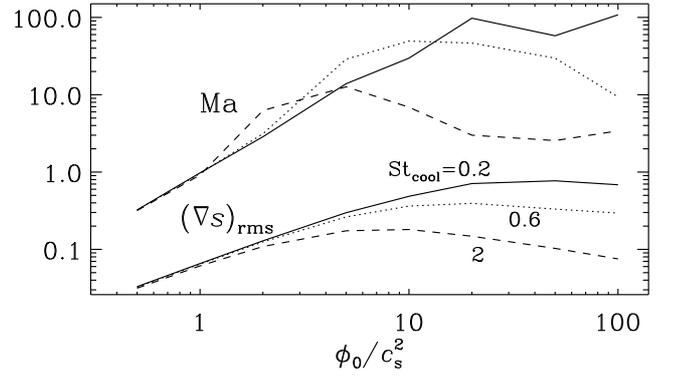}
\end{center}\caption[]{
Dependence of Mach number and the rms value of entropy on $\phi_0$ for $\nu/\cs R=1$
and $\St_{\rm cool}=0.2$, 0.6, and 2 (solid, dotted, and dashed line types, respectively).
}\label{pbaro_nu4_Ma}\end{figure}

\begin{figure*}[t!]\begin{center}
\includegraphics[width=\textwidth]{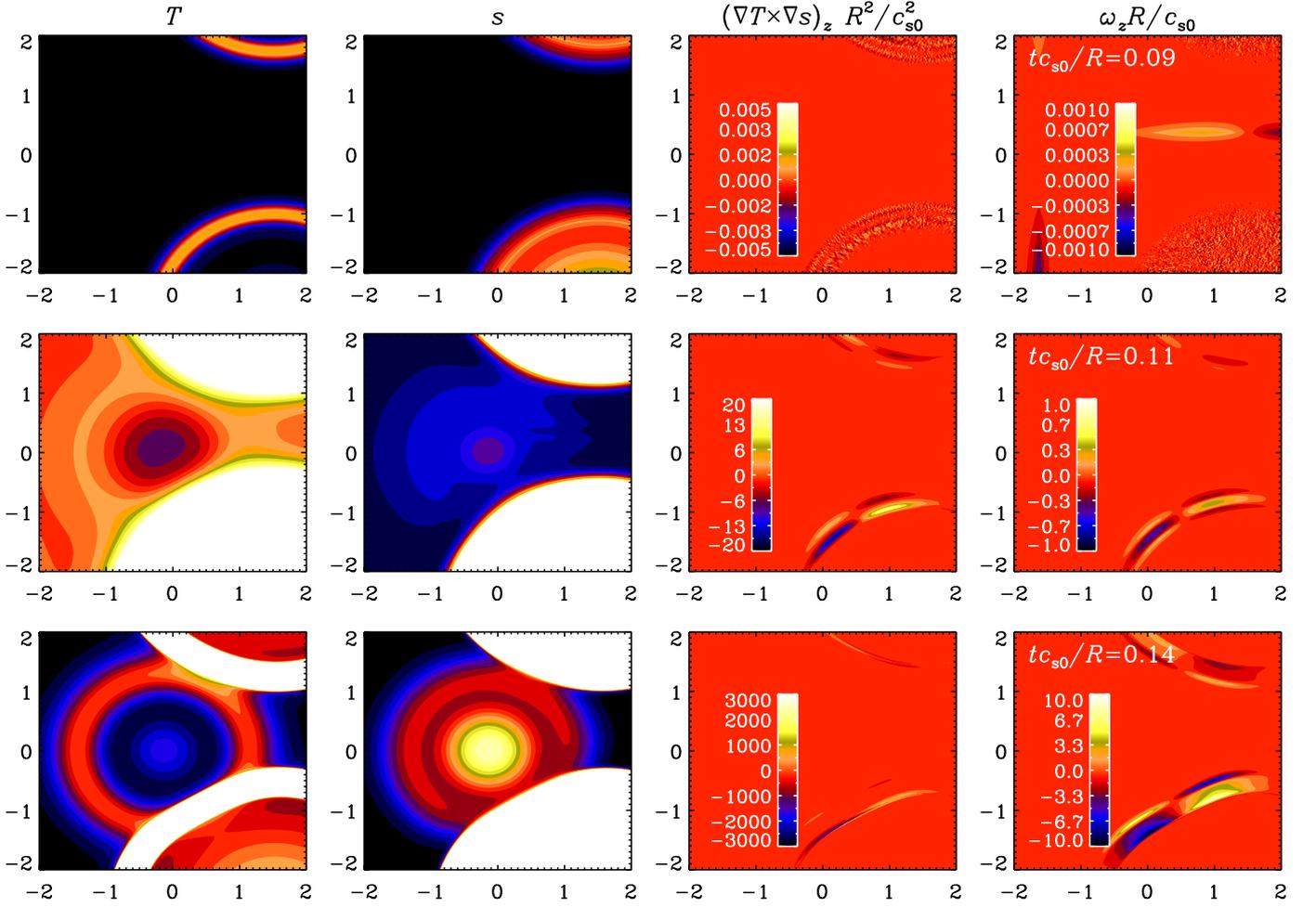}
\end{center}\caption[]{
Images of $T$, $s$, $(\nab T\times\nab s)_z$, and
normalized vertical vorticity for a two-dimensional
run with $\delta t_{\rm force}\csz/R=0.1$ at an instant shortly before
the second expansion wave is launched (top row), and shortly after the
second expansion wave is launched (second and third row).
Note the vorticity production from the baroclinic term in the second
and third row, while in the top row, $(\nab T\times\nab s)_z$ and $\omega_z$
are just at the noise level of the calculation.
Even under our weakly supersonic conditions shock surfaces are well localized and 
the zones of maximum production of vorticity appear to be those in which the fronts
encounter each other.
Here we have used $\phi_0/\csz^2=100$, $\nu=\chi=0.1\csz R$, with
$512^2$ mesh points.
Only the inner part of the domain is shown.
}\label{p2d}\end{figure*}

Finally, we consider the possibility of vorticity generation by the
baroclinic term.
In a preparatory step we study first the dependence of the product
$(\nab T)_{\rm rms}(\nab s)_{\rm rms}$ on both $\Ma$ and $\Rey$
in a one-dimensional model.
In all cases we vary the strength of the forcing amplitude in the range
$1\leq\phi_0/\csz^2\leq500$ for different values of viscosity and
cooling time.
As we increase the value of $\phi_0$, the Reynolds number
increases for a given value of the viscosity.
For small values of $\phi_0$, the Mach number also increases
linearly, where the ratio of $\Ma/\Rey$ increases with increasing
viscosity.
However, for larger values of $\phi_0$ there is saturation and $\Ma$
no longer increase with $\phi_0$.

Furthermore, in the range where $\Ma$ still increases linearly with
$\phi_0$, the rms value of the entropy gradient increases,
but it also saturates when $\Ma$ saturates.
The rms value of the temperature gradient, however, decreases with
increasing values of $\phi_0$, but this seems to be a special
property of the one-dimensional model that is not borne out by the
three-dimensional simulations where it stays approximately constant.

\begin{figure}[t!]\begin{center}
\includegraphics[width=\columnwidth]{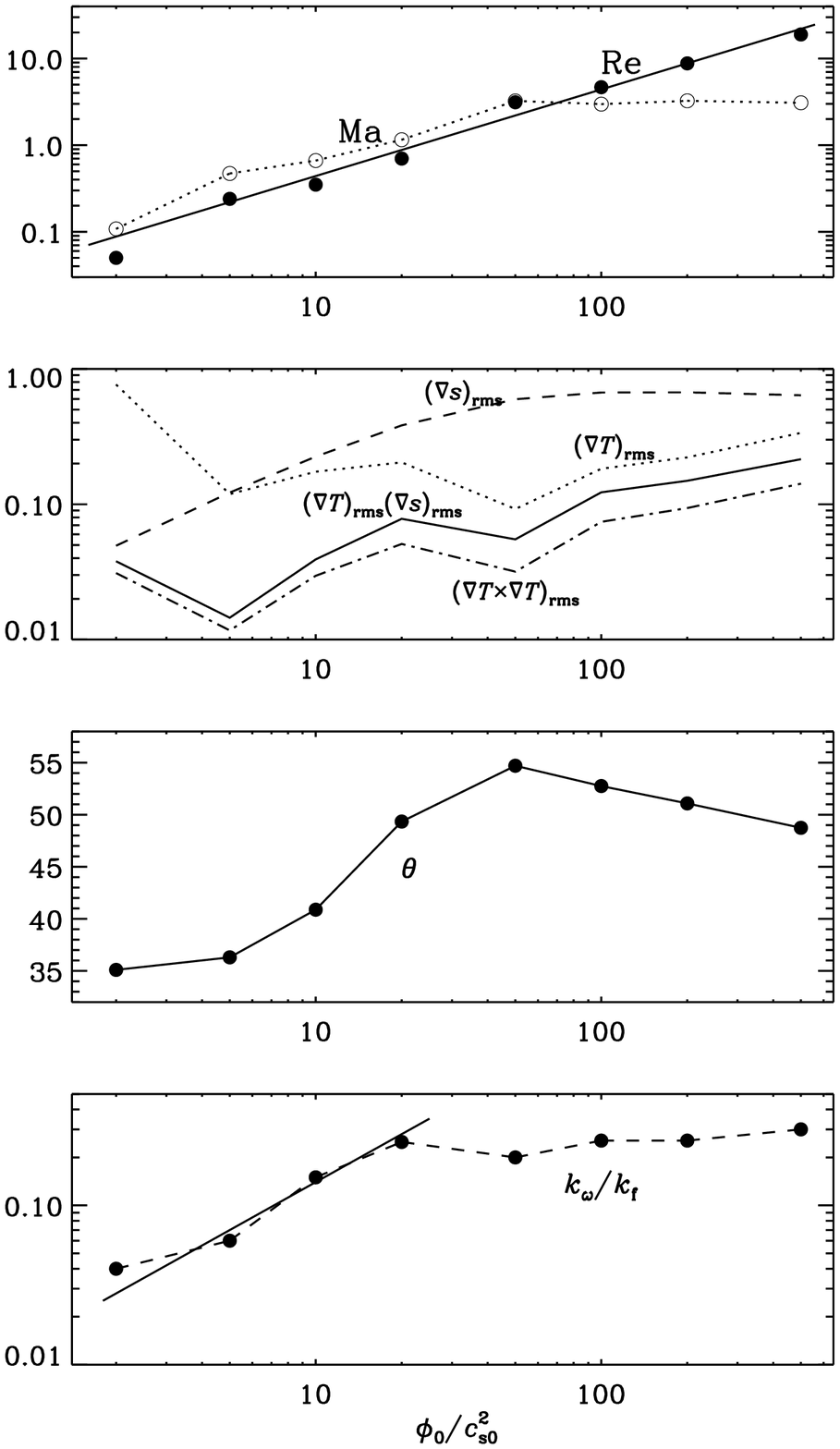}
\end{center}\caption[]{
Dependence of $\Ma$, $\Rey$, the rms values of $\nab s$ and $\nab T$,
the angle $\theta$ between them, as well as $\kom/\kf$,
on $\phi_0/\cs^2$ for $\nu/\cs R=1$.
}\label{pbaro_nu_3d}\end{figure}

Remarkably, the results are fairly independent of the cooling time,
except that the break point where $(\nabla s)_{\rm rms}$ saturates occurs
for smaller values of $\phi_0$ as we increase the cooling time;
see \Fig{pbaro_nu4}.
This break point is also related to the point where the Mach number saturates,
as can be seen from \Fig{pbaro_nu4_Ma}.
However, for longer cooling times there can be longer transients,
making it difficult to obtain good averages.
Therefore we focus, in the rest of this paper, on the case of shorter
cooling times using $\St_{\rm cool}=0.2$.
Another remarkable result is that the normalized value of
$(\nab T)_{\rm rms}(\nab s)_{\rm rms}$ is always of the order
of about $10^{-3}$, independent of resolution, cooling time,
or the value of the viscosity.

Most of the basic features of the one-dimensional model
are also reproduced by two- and three-dimensional calculations.
Two-dimensional simulations have the advantage of being
easily visualized and are therefore best suited for illustrating
vorticity production by the baroclinic term.
In \Fig{p2d} we demonstrate that vorticity production is
associated with the interaction between the fronts of
different expansion waves.
In this example we chose $\delta t_{\rm force}\csz/R=0.1$,
so the first expansion wave is launched at $t=0$ and the
second one at $t\csz/R=0.1$.
The top row of \Fig{p2d} shows that at $t\csz/R=0.09$, i.e.\ just before
launching the second expansion wave, the baroclinic term and the vorticity
are still just at the noise level.
At that time the most pronounced feature is the discontinuity between
the Gaussian expansion waves in the periodic domain.
This leads to negligibly weak vorticity, and no baroclinic term.
At $t\csz/R=0.11$, the effect of the second expansion wave becomes noticeable
in visualizations of both $(\nab T\times\nab s)_z$ and $\omega_z$,
while our visualizations of $T$ and $s$ barely show the second expansion wave.
At $t\csz/R=0.14$, the first expansion wave is clearly no longer circular,
which is obviously associated with the second expansion wave that is now
quite pronounced in our visualizations of both $T$ and $s$.

In order to have a more accurate quantitative determination
of vorticity production, we now consider three-dimensional models.
In \Fig{pbaro_nu_3d} we show the dependence
of various quantities on $\phi_0$ for $\St_{\rm cool}=0.2$
and $\nu/\cs R=1$.
In all cases we use $128^3$ mesh points and average the results over
between 20 and 70 turnover times.
Note that here $\Rey\approx0.05\,\phi_0/\csz^2$.
Given that Re depends inverse proportionally on $\nu/R\csz$, we can
also write $\Rey\approx0.05\,\phi_0 R/\csz\nu$.
The Mach number saturates at about $\Ma=3$, and the rms value of
the entropy gradient increases up until this point.
Given that the rms value of the temperature gradient also stays
approximately constant, we find a weak increase of
$(\nab T)_{\rm rms}(\nab s)_{\rm rms}$.
The value of $(\nab T\times\nab s)_{\rm rms}$ is always found to be a
certain fraction below this value, resulting in a typical baroclinic angle
of about 45 degrees; see the third panel of \Fig{pbaro_nu_3d}.
Finally, the amount of vorticity production in terms of $\kom/\kf$
is about 0.3 for $\phi_0/\csz^2\ga20$.
For smaller values, on the other hand, there is an approximately
linear increase with $\kom/\kf\approx0.014\,\phi_0/\csz^2$.

\begin{figure}[t!]\begin{center}
\includegraphics[width=\columnwidth]{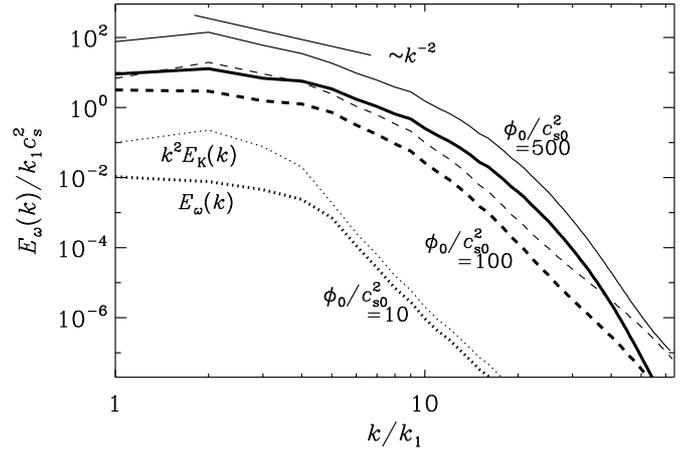}
\end{center}\caption[]{
Time-averaged enstrophy spectra, $E_\omega(k)$ (thick lines), compared
with $k^2E_{\rm K}(k)$ (thin lines below the corresponding thick lines),
for the three-dimensional baroclinic case with
$\phi_0/\csz^2=10$ (dashed), 100 (dotted), and 500 (solid lines).
The $k^{-2}$ slope is shown for comparison.
In all three cases we have $\kf/k_1=4$.
}\label{pspec_comp_enstr_baro}\end{figure}

The possibility of spurious vorticity is easily eliminated in this case
by looking at enstrophy spectra; see \Fig{pspec_comp_enstr_baro}, where
we compare $E_\omega(k)$ with $k^2E_{\rm K}(k)$.
All spectra fall off rapidly with increasing $k$.
Thus, even though the initial vorticity generation occurred evidently
at the smallest available scales, once the flow becomes fully developed,
most of the enstrophy resides at scales equal to or larger than the
driving scale.
Furthermore, the spectra of $E_\omega(k)$ and $k^2E_{\rm K}(k)$ are close
together, suggesting that the vorticity is close to its maximal value.

\section{Applications}

The level of vorticity that is produced in the usual case of solenoidal forcing
of the turbulence is such that $\kom/\kf\approx1$ \citep[see, e.g.,][]{B01}.
For turbulence whose driving force has finite correlation time
($\St_{\rm  force}=0.3$, for example), and small values of $\Rey$,
we have $\kom/\kf=O(1)$ when $\Co\ga10$.
However, for larger values of $\Rey$, the turbulence becomes vortical
already for smaller values of $\Co$.
Comparing with the galaxy, we have $\Omega\approx10^{-15}\s^{-1}$,
$\urms=10\kms$, and an estimated correlation length of about $70\pc$,
so $\kf=3\times10^{-20}\cm$, so $\Co=0.07$, which is rather small.
Thus, rotation may not be able to produce sufficient levels of vorticity.
Given that in galaxies with flat rotation curves, $S\approx-\Omega$,
shear should not be very efficient either.
However, the Mach numbers are undoubtedly larger than unity in the
interstellar medium, so this should lead to values of $\kom/\kf\approx0.3$,
which is the saturation value found in \Fig{pbaro_nu_3d}.
Given that one of the reasons for studying the production of vorticity
is the question of dynamo action, we should point out that such values
of $\kom/\kf$ are large enough for the small-scale dynamo.
Large-scale dynamo action should be possible in galaxies as well,
because of their large length scales, but it suffers from the well-known problem
of a small growth rate. It then remains difficult to explain 
large-scale magnetic fields in very young galaxies \citep{Beck96}.

The question of vorticity generation is also important in studies of
the very early Universe, where phase transition bubbles are believed to be
generated in connection with the electroweak phase transition
\citep{KK86,Ignatius94}.
Here the equation of state is that of a relativistic fluid,
$p=\rho c^2/3$, where $c$ is the speed of light.
Thus, there is no baroclinic term and no obvious source of vorticity.
However, the relativistic equation of state may be modified at small
length scales, but it is not clear that this can facilitate significant
vorticity production.

\section{Conclusions}

The present work has demonstrated that vorticity production is actually
quite ubiquitous once there is rotation, shear, or baroclinicity.
This implies that the assumption of potential flows as a model for
interstellar turbulence might be of academic interest and could only
be realized under special conditions of weak forcing, weak rotation,
and no shear.
In galaxies, however, the shear and Coriolis number are well
below unity, leaving only the baroclinic term as a viable candidate
for the production of vorticity.
This agrees with early work of \cite{K99b}, who analyzed the production
terms in supersonic, supernova-driven turbulence quantitatively;
see also \cite{GLM97}, who showed that on long enough time scales
significant vorticity can also be produced for subsonic flows.
We have also observed how vorticity is mainly produced close to shock front encounters.
This motivates a more detailed investigation of these zones as the next step in the study
of vorticity generation in the interstellar medium.
It should also be pointed out that the baroclinic term corresponds to
the battery term in the induction equation \cite{Kuls97,BS05}.
Thus, when studying the possibility of dynamo action, this battery
term provides an intrinsic and well defined seed for the dynamo
and should therefore be included as well.

\begin{acknowledgements}
We thank an anonymous referee for making a number of useful suggestions
that have improved our paper.
We acknowledge the allocation of computing resources provided by the
Swedish National Allocations Committee at the Center for
Parallel Computers at the Royal Institute of Technology in
Stockholm and the National Supercomputer Centers in Link\"oping.
This work was supported in part by
the European Research Council under the AstroDyn Research Project 227952
and the Swedish Research Council grant 621-2007-4064.
\end{acknowledgements}

\vfill\bigskip\noindent\tiny\begin{verbatim}
Header: /var/cvs/brandenb/tex/joern/spherical/paper.tex,v 1.34 2011-02-25 14:33:39 brandenb Exp $
\end{verbatim}

\end{document}